# Features of surface structuring of direct and indirect band gap semiconductors by femtosecond laser

**Berezovska N.I.[1], Dmitruk I.M.[1], Hrabovskyi Ye.S.[1], Kolodka R.S.[1], Stanovyi O.P.[1], Dmytruk A.M.[2], Blonskyi I.V.[2]**

[1] Faculty of Physics, Taras Shevchenko National University of Kyiv. Volodymyrska Street, 64/13, Kyiv-01601, Ukraine.
e-mail: n_berezovska@univ.kiev.ua, +38(067)4472540;
igor_dmitruk@univ.kiev.ua, +38(066)7752127;
hrabovskye@gmail.com, +38(095)0834192;
rkolodka@gmail.com, +38(097)3915689;
astanovyi@gmail.com, +38(099)1100429;

[2] Photon Processes Department, Institute of Physics of NAS of Ukraine. Prospect Nauky, 46, Kyiv-03028, Ukraine.
e-mail: admytruk@iop.kiev.ua, +38(044)5251543; blon@iop.kiev.ua, +38(044)5259810

**Abstract** The impact of femtosecond (fs) laser radiation on semiconductors with direct (ZnSe, GaAs, CdZnTe) band gap, with the structurally induced direct-to-indirect band gap transition ($PbI_2$, GaSe) and indirect band gap (Si) has been studied. The fs-laser treatment of semiconductors has been performed in the multi-pulse regime in air environment. The influence of fs-laser radiation parameters on surface morphology of the semiconductors has been analysed by scanning electron microscopy (SEM) and 2D Fourier transform of SEM images, optical photoluminescence spectroscopy. Under the treatment with the fundamental fs-laser radiation (800 nm, about 130-150 fs) both low spatial frequency LIPSS and high spatial frequency LIPSS have been observed. Specific features of LIPSS of two types (low spatial frequency LIPSS and high spatial frequency LIPSS) and other structure peculiarities (grooves, grains, different defects in periodic structure, i.e. loops, strip breaks, and ablation products) have been also analysed. The formation mechanisms of LIPSS are also considered within the scope of two approaches, namely an electromagnetic approach and matter reorganization processes.

## 1. Introduction

In recent years, promising applications of micro- and nano-textured surfaces in different areas of science and technology stipulate the active study of the surface modification with different methods that results in the formation of the surface micro- and nanostructures. In contrast to such complex techniques, as, for example, lithographical techniques, ultra-short laser processing could create in one-step regime 3D periodic structures at the surface of various materials, such as metals [1–3], semiconductors [4–7] and dielectrics [8–10]. The interest to the formation of laser-induced periodic surface structures (LIPSS) or ripples is constantly growing since 1965, when Birnbaum M. [11] demonstrated the occurrence of the periodic structures at the surface of single-crystal specimens of high-purity semiconductors of germanium, silicon, GaAs, GaSb, InSb, and InAs after an irradiation by a ruby laser. The method of direct femtosecond laser structuring allows to treat a relatively large area of the sample with controllability and reproducibility of the processing results. Practical benefits of LIPSS at semiconductor surfaces cover wide range of applications in photonics, plasmonics, optoelectronics, biochemical sensing, wettability changes, etc. Thus, the texturing of the surface of photovoltaic elements belongs to a standard treatment procedure aimed at reducing the reflection losses [12, 13], increasing the robustness against defects and perturbations [14]. The combination of semiconductor surface relief with metal features induces the efficient light trapping and an increase of the absorption in thin layers due to the involvement of the surface plasmons-polaritons (SPPs) [15–17]. Formation of LIPSS on semiconductors and subsequent plating with gold or silver nanofeatures can be used for the producing of the substrates for surface-enhanced Raman scattering (SERS) [18, 19].

In general, both the fs-laser beam parameters and material properties determine the results of the light–matter interaction. The classification of the LIPSS could be found for example, in [20, 21]. LIPSS with the structure period close to the laser wavelength ($\Lambda \cong \lambda$) are called low spatial frequency LIPSS (LSFL). Commonly their orientation is perpendicular to the laser beam polarization. Usually this type of ripples is inherent to the strong absorbing materials, such as semiconductors and metals. In some cases, LIPSSs manifest direction parallel to the laser beam polarization, and frequently their period is larger than the laser wavelength. Such structures are also called grooves. Another type of LIPSS is called high spatial frequency LIPSS (HSFL). Such LIPSS are characterised with spatial periods significantly smaller than the laser wavelength ($\Lambda < \lambda/2$). They are often revealed for the weakly absorbent materials such as dielectrics. HSFL are also observed for the semiconductor materials. HSFL can exhibit both orientations (parallel or perpendicular to the laser beam polarization). Different types of LIPSS are associated with different formation mechanisms.

Along with accelerating experimental research activity on LIPSS formation on different materials, theories and numerical methods concerning origin of LIPSS are actively developed. Leading scientists who intensively study LIPSS formation divide the approaches to the LIPSS creation into two types, i.e. electromagnetic approach describing the impact of optical energy into the solid and matter reorganization approach describing the redistribution of surface near the area of laser-excited material [21]. Within electromagnetic theory on the formation of LIPSS the incident light scatters at the roughness of the sample surface exciting surface electromagnetic waves (SEWs) including surface plasmon polaritons (SPP). LIPSS formation usually is a result of multi-pulse laser treatment in the ablative regime. In addition to the interference between the incident laser beam and SEWs (i.e. SPPs) additional second-order contributions can be caused by the interference of counterpropagating plasmon waves [22]. This approach can be applied to the initially plasmonically nonactive materials, such as semiconductors and dielectrics, which can be turned into a metallic state at an excess of a critical density of electrons in the conduction band during fs-laser irradiation [5, 23–25]. Thus, Bonse and co-workers [23] combined the Sipe's theory with a Drude model to quantify the change of dielectric permittivity as a function of the laser-induced electron density $N_e$ in the conduction band of silicon. Furthermore, after the initial formation of the periodic profiles it should be also taken into account the subsequent coupling of the electronic system and the lattice of the solid through electron–phonon coupling, as well as thermal diffusion, electron diffusion and phase transitions of melting substance [26, 27].

Regardless the progress of elucidation of the LIPSS origin further investigations of the nature of LIPSS in different materials and hybrid structures are of great importance [28].

We present a survey of the impact of femtosecond laser impulses on set of semiconductors with direct and indirect band gap (GaAs, ZnSe, GaSe, $PbI_2$, CdZnTe and Si), compare the peculiarities of morphology of textured surfaces obtained in similar fs-laser processing setup, and discuss the nature of these patterns, as well as spectroscopic manifestation of the effect of fs-laser treatment on optical properties of corresponding semiconductor.

## 2. Experimental details

Micro- and nanostructuring of semiconductors under study have been carried out with a femtosecond laser, namely a Ti-sapphire laser system which consists of a Mira-900F femtosecond oscillator with a Legend-HE chirped pulse amplifier (both from Coherent, USA). The main characteristics of the laser have been the following: central wavelength of 800 nm, a pulse duration of 130 - 150 fs, a pulse repetition rate of 1 kHz, a pulse energy about 0.8 - 1 mJ. Cylindrical lenses and spherical lens (in the case of GaAs texturing) are used for the beam focusing onto the surface sample. A typical power density at the sample surface is of the order of $10^{12}$ W/cm$^2$. A laser power meter "Field Master GS" with a detector head "LM-10" (Coherent, USA) was used for the measurements of the average laser power. In some experiments, the diaphragm for the incident beam has been used, and its presence in the setup was taken into account while the pulse irradiation energy density was evaluated. The surface texturing has been realized in scanning beam regime at a velocity range of 0.45…1 mm/s. During the laser processing, a vertically standing sample stage moves at a defined velocity. The horizontally polarized laser beam is incident normally onto the sample surface. The laser processing has been performed in air environment in the multi-pulse regime.

The influence of fs-laser radiation parameters on surface morphology of the samples has been analyzed by scanning electron microscopy (SEM) with AURA 100 SEM (SERON Technology Inc.) and 2D Fourier transform (2D-FFT) of SEM images. 2D Fourier transform has been performed using Gwyddion 2.60 software. We also present the resultant 2D-FFT in 3D view for the better visualization of its shape. Sometimes 3D view presentation has been rotated relative to the orientation of the corresponding SEM image and its 2D-FFT presentation. It should be noted that for a more convenient presentation of 2D-FFT results, we show its profile in a certain direction in µm$^{-1}$ (end bars represent the signal summation width in specified direction), and its low-frequency corrected variant in nanometers, where spatial periods of the structure are clearly seen. This graph of the profile could be further analyzed by the peak deconvolution.

The photoluminescence (PL) spectra have been measured with the setup based on a single-grating spectrometer MDR-3 (LOMO) under an excitation of semiconductor laser $\lambda_{ex} = 406$ nm ($E_{ex} = 3.05$ eV), $\lambda_{ex} = 532$ nm ($E_{ex} = 2.3$ eV) and pulsed nitrogen laser $\lambda_{ex} = 337$ nm ($E_{ex} = 3.68$ eV).

# 3. Peculiarities of femtosecond light-semiconductor interaction in the case of direct band gap semiconductors

***Direct band gap semiconductor zinc selenide (ZnSe)*** belongs to the popular optoelectronic materials. ZnSe is used in light-emitting diodes, photodetectors, X-ray detectors, field emitters, elements of solar cells and other devices. It has wide band gap – $E_g = 2.7$ eV at 300 K. There are only a few papers describing the fs-laser treatment of ZnSe [29–35]. As a rule, LIPSSs with periods around 160 – 180 nm have been observed.

We also studied fs-laser processing of ZnSe single crystal with maximum resistivity ($\rho \geq 10^{12}$ Ω cm) and a minimum concentration of impurities using linearly polarized Ti/sapphire fs-laser [36]. The fs-laser pulses with the central wavelength of 800 nm with the energy density of around 0.04-0.05 J/cm², the effective number of laser pulses varied in the range of ~300-400, the pulse duration of 140 fs at a repetition rate of 1 kHz have been applied for the ZnSe texturing. As impurities and native defects could affect electronic and optical properties of ZnSe, the influence of the fs-laser processing on luminescent properties of ZnSe has been investigated to elucidate the changes in the quality of ZnSe single crystals after fs-laser processing. During fs-laser processing at the regime of the lower fluence LIPSSs of two periods 220 and less pronounced 640 nm were revealed at the surface (see Fig. 1). The periods values were defined by 2D Fourier transform of SEM images Fig. 1, a, represents the central and peripheral fields of the fs-laser beam action on the sample surface. At Fig. 1, b, the incipient periodic structures are seen. These structures are forming along the existing cracks at the sample surface or located randomly. At the same time the obtained structures trace the laser beam polarization, namely these structures are oriented perpendicular to laser beam polarization. Fig. 1, c, represents the central part of the area processed with the fs-laser beam, d – its Fourier transform. LIPSS are rather homogeneous along the surface.

The occurrence of the LSFL and HSFL at the fs-laser treated ZnSe surface indicates the complicated nature of the process of fs-laser influence at the semiconductor surface. The laser-induced changes of ZnSe carrier concentrations specify periods of obtained LIPSS. Authors of [23, 29, 35] emphasize that transient changes in material properties under laser irradiation are of crucial importance in the LIPSS formation. The energy of incident light 800 nm (1.55 eV) is less than the band gap of ZnSe single crystal, so the electrons in the valence band will be excited up to conduction band via two-photon absorption when the sample is irradiated by femtosecond laser pulses. Therefore, the complex refractive index of the laser treated ZnSe is a function of the carrier density of conduction band electrons.

LSFLs are characterised by the periods smaller than the laser wavelength, since the surface damage begins at carrier concentrations close to the critical carrier density of surface plasmon polaritons (SPP) excitation [23]. For semiconductors it appears in a relatively narrow fluence range.

The period of HSFL in transparent materials usually follows the empirical formula $\Lambda \simeq \lambda/2n$, where n is the refractive index of an untreated material and λ is the wavelength of the incident fs-laser beam. We obtained slightly higher periods of LIPSS (from 182 to 214 nm depending on a slight change in processing parameters for corresponding ZnSe surface). The modified fs-laser-excited refractive index n* for ZnSe is lower than that for untreated material [29, 33]. Thus, the periods of HSFL obtained in our experiments are in good accordance with the periodicity around 200 nm for bulk ZnSe demonstrated in other papers [33, 34].

For the evaluation of the possibility of further perspective applications of textured semiconductors, in particular ZnSe, the optical properties of modified surface should be examined. We studied near-band-edge emission and emission related to the impurities or structural defects of untreated and fs-laser treated ZnSe single crystal at room and liquid nitrogen temperatures (see Fig. 2). PL bands in the region from 500 to 700 nm determined by intrinsic point defects are of the same order of magnitude for untreated and textured areas of ZnSe. In general, it should be stated that the high structural perfection of ZnSe single crystal is not affected significantly under fs-laser processing at the levels of the energy density of around 0.04-0.05 J/cm$^2$, that do not exceed an ablation threshold for ZnSe which is about 0.7 J/cm$^2$ [37].

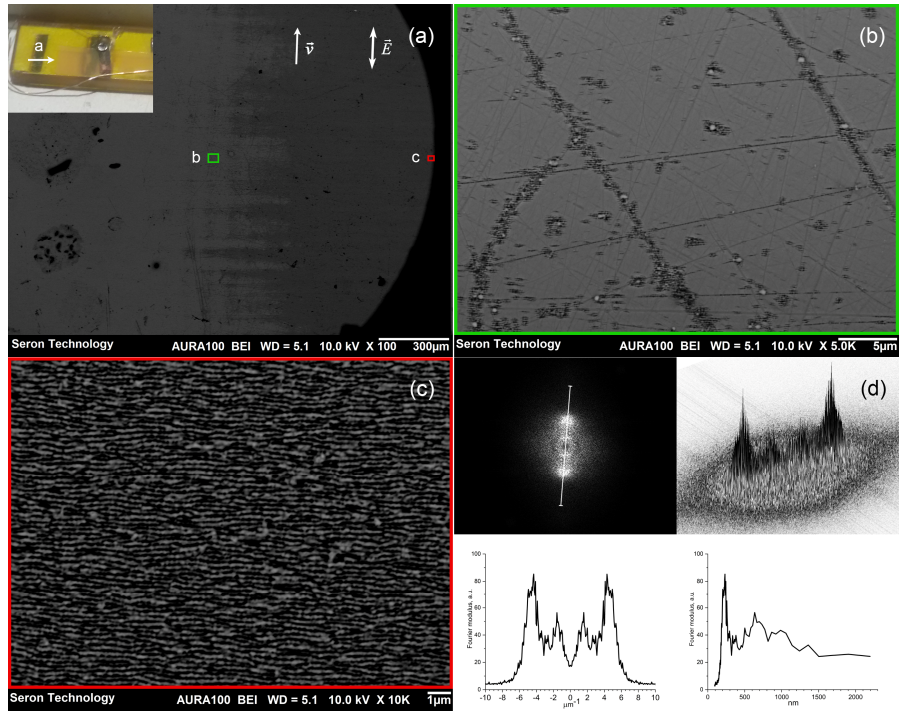

**Fig. 1** SEM images of ZnSe surface treated with fs-laser beam focused with cylindrical lens (a, b, c): (a) general view of the treated sample, (b) surface irradiated by the beam periphery part, (c) central part of structured area; (d) 2D Fourier transforms of a SEM image (c) in two views; lower part of (d) presents the Fourier transform profile along the line at 2D Fourier transform and the Fourier transform profile in nanometers. The inset in (a) presents the view of fs-laser treated ZnSe single crystal

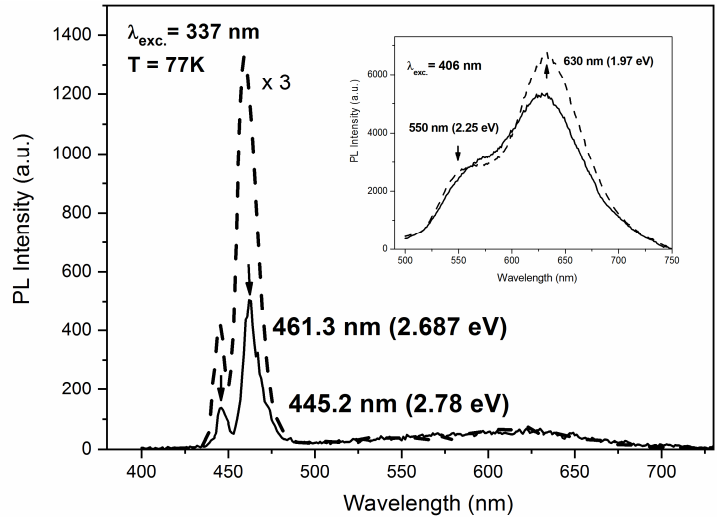

**Fig. 2** PL spectra of fs-laser treated (solid line) and untreated (dashed line) surface of ZnSe sample at 77 K under an excitation of 337 nm and 406 nm (inset)

## *Direct band gap semiconductor gallium arsenide (GaAs)*

($E_g$ = 1.42eV at 300 K) belongs to A3B5 semiconductors. GaAs characterized with a higher electron mobility and higher thermal stability than silicon is perspective for the application in multi-junction or cascade solar cells, micro-optical components, and terahertz emitters. Production of functional surface frequently require the texturing of GaAs surface. Femtosecond-laser modification of GaAs has been reported in a number of papers [4, 38–40].

In our experiments, GaAs wafers were treated with linearly polarized fs-pulses (wavelength of 800 nm, the pulse energy density of ~0.4 J/cm²). Observed LIPSS are orientated perpendicular to the laser beam polarization with period of 725 nm, additional feature in FFT (360 nm) corresponds to splitting of LSFL ridges. This process is sometimes associated with formation of some types of HSFL. There are several proposed mechanisms, namely enhancement of local electric field by surface morphology [41] or laser induced point defect accumulation and diffusion [38]. Supra-wavelength-sized quasi-gratings oriented parallel to the laser beam polarization, although known as grooves [42] with period of ~1.7 μm (see Fig. 3, and Fig. 4) although can be seen. Their formation is associated with heat accumulation and hydrodynamic effects [6]. The origin of the LSFL attributed to the electromagnetic approach, in particular the interference between the incident laser beam and SEWs including SPPs.

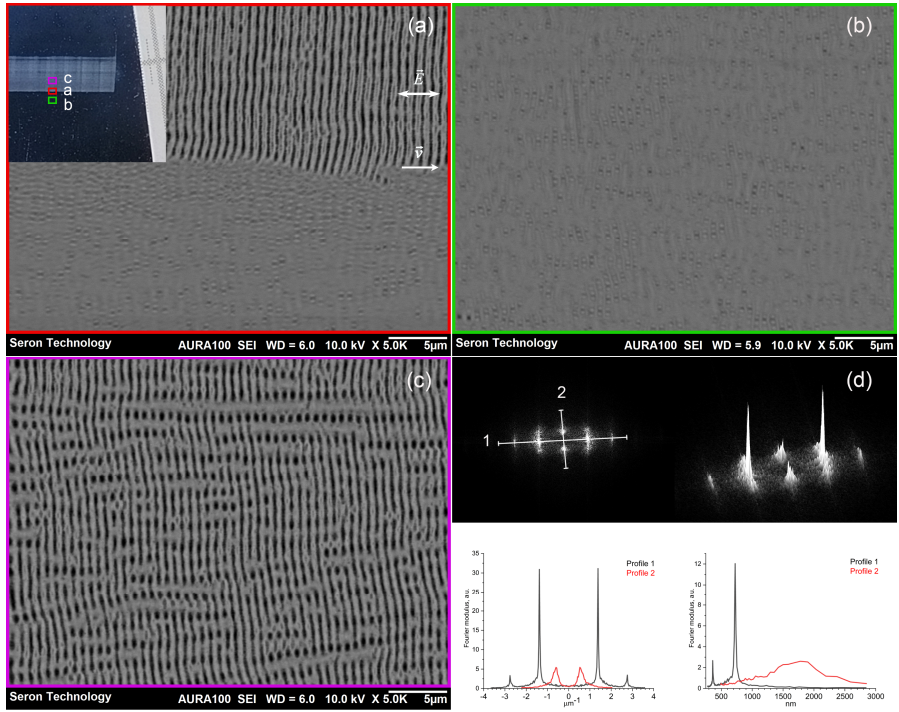

**Fig. 3** SEM images of edge, outlying and central parts of an area scanned by laser beam focused with cylindrical lens at GaAs surface (a, b, c); 2D Fourier transform of a SEM image (c) of fs-laser treated GaAs surface (d); lower part of (d) presents the Fourier transform profiles along the lines 1, 2 in 2D Fourier transform (d) and the Fourier transform profiles in nanometers. An inset presents view of the fs-laser beam scanning area at GaAs surface

Cylindrical lens provides more smooth spatial variation of irradiation parameters athwart structured line. In combination with diaphragms of different shape, this allows to create large areas of homogeneous structures in a single scan, or one can study effects of different fluence and effective pulse number in different parts of the scan.

Fig. 4 shows the same sample irradiated with beam focused by spherical lens. Resulting structures are less regular, structures on beam periphery are quite different (Fig. 3 b and 4 b).

The laser fluence change results in change of morphology, namely appearance of porous and grain structures, periodicity violation, etc. (see Fig. 4, c, d). The topology of the surface undergo drastic changes under increased fluence – the grain-porous structure is forming. On the grains with characteristic size of 1.7 µm LSFL with period around 500-800 nm are formed. 2D Fourier transform of corresponding SEM image demonstrates the existence of two structural features in two perpendicular directions. The origin of such structure may be related to the effects of matter reorganization, in particular hydrodynamic effects of the transiently melted surface, material instabilities or defect creation, diffusion, or erosion effects enforced by ultrafast laser processing.

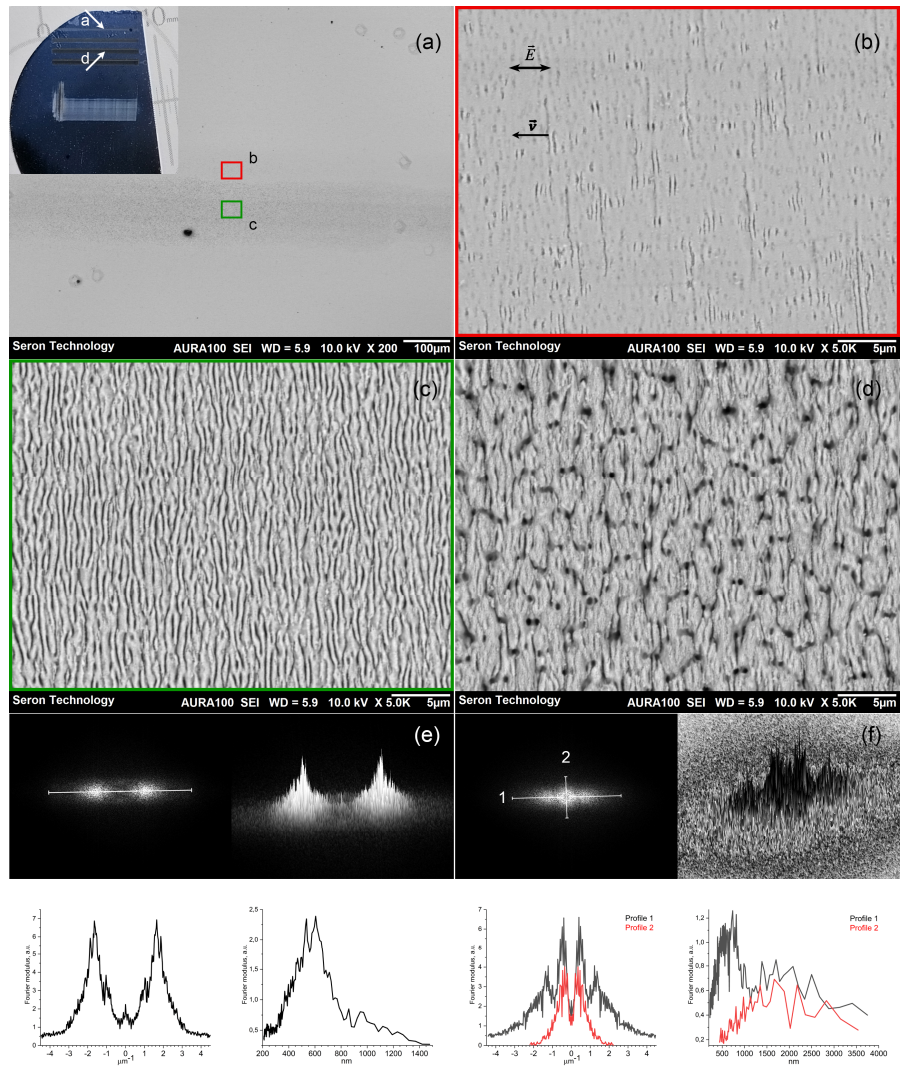

**Fig. 4** SEM images of GaAs surface structured by scanning with spherical lens focused laser beam. (a) general view of structured area, (b, c) central and outlying parts of (a); (d) central part of area structured with higher fluence; (e) 2D Fourier transform of a SEM image (c); (f) 2D Fourier transform of a SEM image (d); lower part of the figure presents the Fourier transform profile along the line at 2D Fourier transform (e) and the Fourier transform profiles along lines 1, 2 at 2D Fourier transform (f), and corresponding Fourier transform profiles in nanometers. An inset in (a) presents view of the fs-laser beam scanning areas at GaAs surface

Obtained porous structure at the GaAs surface after the fs-laser treatment with higher power density (Fig. 4, d) resembles chemically etched GaAs surface. It is characterized with rather high homogeneity.

To evaluate possible changes in optical properties the near-band-edge emission of untreated and fs-laser treated GaAs samples represented at Fig. 3 has been studied at liquid nitrogen temperatures under an excitation of 532 nm (see Fig. 5). PL band at around 1.51 eV caused by the recombination of free excitons has been observed for untreated and fs-laser treated surfaces at T= 77K. The intensity of PL band at fs-laser treated GaAs surface 2.5 times smaller than the one for untreated GaAs surface.

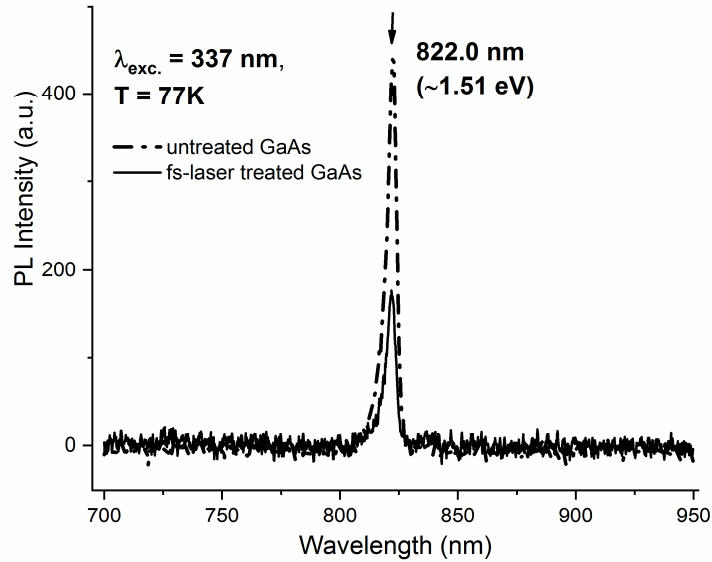

**Fig. 5** PL spectra of fs-laser treated (solid line) and untreated (dashed line) surface of GaAs sample at 77 K under an excitation of 532 nm

***Direct band gap semiconductor cadmium zinc telluride* (CdZnTe) or CZT** is an alloy of zinc telluride and cadmium telluride. CdZnTe is a direct band gap semiconductor with the band gap that varies from approximately 1.4 to 2.2 eV depending on Zn concentration. It possesses a high electro-optic coefficient and transparency in the mid-infrared region. CdZnTe is perspective for some applications, including semiconductor radiation detectors, photorefractive gratings, electro-optic modulators, solar cells, and terahertz detectors. Graded CdZnTe was proposed as an alternative for the pure CdTe in thin film solar cells for the increasing their efficiency [43].

Most of the experiments on laser processing of CdZnTe have been carried out with nanosecond (ns) laser pulses [43–47]. Nivas J. and co-authors [48] stated that the fs laser induced surface processing of CdZnTe has not been investigated yet. They carried out an experimental study of laser ablation and surface structuring of CdZnTe in air. The fs-laser pulses with the central wavelength around 800 nm, a pulse duration of ≈35 fs have been used for three sets of the laser processing conditions (from the low peak fluence of 0.64 J/cm$^2$ up to the highest peak fluence of ≈7.0 J/cm$^2$). Both LSFL and HSFL perpendicular to the laser beam polarization have been observed. These LIPSSs are attributed to the progressive aggregation of randomly distributed nanoparticles, accompanied by the process of laser ablation at high pulse energy that produces a deep crater.

In our experiments on fs-laser treatment of CdZnTe the linearly polarized fs-laser radiation with central wavelength of 800 nm, the pulse energy density of ~0.4 J/cm$^2$ has been used. Slight change of fluence was attained by the sample shift from the focusing lens. The HSFLs with period of 220 nm are seen at Fig. 6, c, e. It should be pointed out at the observed changes of the morphology at different edge areas of the laser beam scanning line at CdZnTe surface (see Fig. 6, d, g, h). At the part of the laser beam scanning line the clearly defined LSFL with the period of 550 nm has been formed. On Fig. 6 g, we can see that LSFL area exists in a strip near the edge of structured area and the structure become dominated by grooves closer to the middle. HSFL comprised of aggregated nanoparticles can be obtained not only around the crater but although by prolonged irradiation by low intensity beam as in our case. Less pronounced outlying structure on the upper part of b as well as structure in corresponding area for GaAs (Fig. 3 b) needs further investigation.

The specific view of the scanned area (Fig. 6 b) is due to the fact that scanning on short distances with subsequent change in the beam intensity was performed. Therefore, the scanned strip consists of places where the beam stopped and stood, and areas where the scanning took place. The technical features of mechanics lead to the accelerated movement at speed-up and braking mode that results in specific view of the structured area.

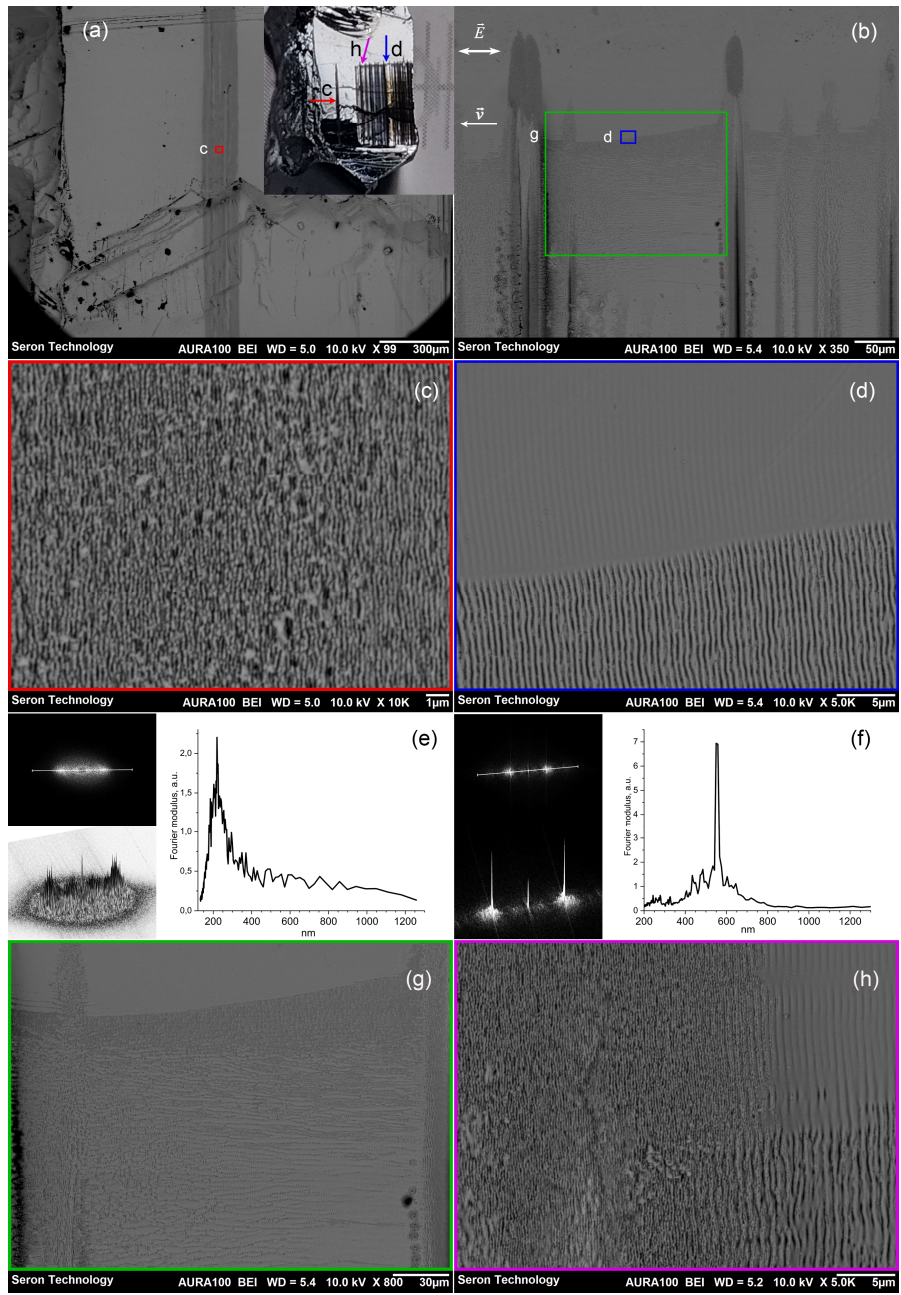

**Fig. 6** SEM image of stationary fs-laser treated area at CdZnTe surface (a); SEM images of different edge laser beam scanning areas at CdZnTe surface (b, d); SEM image of the central part of the area structered by the stationary beam (c); 2D Fourier transform of a SEM image (c) pre-

sented in two views and the Fourier transform profile along the line at 2D Fourier transform presented in nanometers (e); 2D Fourier transform of a SEM image (d) in two views and corresponding Fourier transform profile along the line at 2D Fourier transform presented in nanometers (f); SEM image of the laser beam scanning area (g); SEM image of the place where adjoin areas formed by the stationary beam and by the scanning beam (h). An inset in (a) presents the view of the fs-laser treated areas at CdZnTe surface

## 4. Femtosecond laser processing of the semiconductors with the transition of band gap from a direct type to an indirect type

***Layered semiconductor lead iodide (PbI₂)*** belongs to the transition metal halides (TMH) family applicable for perovskite solar cells, photodetectors, nuclear radiation detectors. This semiconductor is characterized with specific structure and interesting optical properties. Direct band gap of 2.26 eV in bulk $PbI_2$ transforms to an indirect band gap with an energy of 2.64 eV for a single-layer $PbI_2$ [49]. This thickness dependent band gap determines the photoluminescence and optoelectronic properties of $PbI_2$.

In the literature, we did not find any scientific articles directly related to fs-laser processing of $PbI_2$. There were attempts of laser patterning of the solar cell layers of perovskite solar modules by nanosecond laser pulses with durations of about $\approx$ 30 ns at a wavelength of 532 nm [50]. Authors observed LIPSS with residuals composed of $PbI_2$ remains even at higher fluences.

In our experiments, we carried out the direct fs-laser texturing of $PbI_2$ surface with the linearly polarized fs-laser radiation with central wavelength of 800 nm, the pulse energy density of ~0.4-0.5 J/cm². The results of morphology analysis of untreated and fs-laser treated surfaces of $PbI_2$ are presented at Fig. 7. We discovered the exfoliation of the $PbI_2$ layers during fs-laser treatment (see Fig. 7, b). We observed the beginning of the LSFL formation with a period of around 700 nm at applied fluences (see Fig. 7, c, d). The obtained LSFLs are irregular and weakly pronounced.

***Layered semiconductor gallium selenide (GaSe)*** is a layered semiconductor with an indirect band gap of ~2.1 eV. The thickness of GaSe films has a strong influence on its electronic structure. 2D GaSe thin films have a larger band gap than the bulk GaSe. A single layer consists of covalently bonded Se-Ga-Ga-Se atoms. Each Ga atom is tetragonally-oriented to three Se atoms and one Ga atom. The layers are stacked together via van der Waals interaction. According to theoretical calculations [51–53] bulk GaSe has a direct band gap. A direct-to-indirect band gap transition occurs as the number of layers is decreasing below some critical value. As the number of layers decreases to below 7, the calculations show that the conduction band minimum remains at the Brillouin zone center (Γ point), but the valence band maximum shifts away from Γ point to nearby k points [52]. The energy band in the vicinity of the Γ point exhibits a Mexican-hat-like dispersion.

GaSe crystals are characterised with wide optical transparency of 0.65 to 18 μm. GaSe has been used as a photoconductor, a second harmonic generation crystal in non-linear optics. GaSe is used to generate THz radiation up to 31 THz and above. GaSe also is perspective for applications in solar cells. According to available literature data, there are only few articles devoted to GaSe laser processing [54]. There is lack of information about fs-laser processing of GaSe semiconductor.

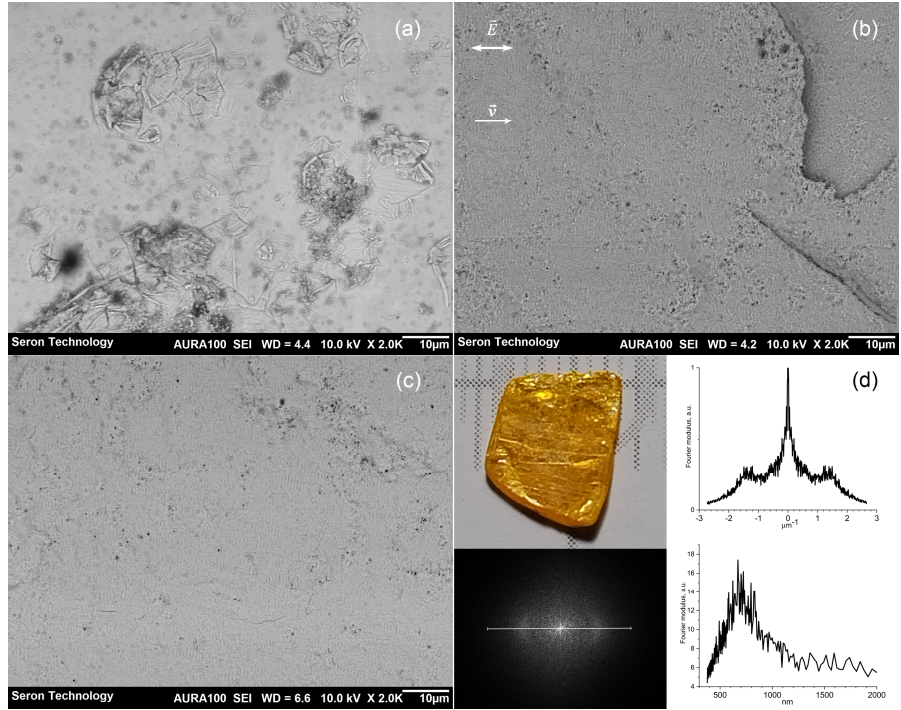

**Fig. 7** SEM image of untreated PbI$_2$ surface (a); SEM images of different fs-laser treated areas of PbI$_2$ surface (b, c); (d) represents the view of PbI$_2$ sample, 2D Fourier transform of a SEM image (c), the Fourier transform profile along the line in 2D Fourier transform and corresponding Fourier transform profile in nanometers

After fs-laser treatment at the pulse energy density of ~0.5-0.6 J/cm$^2$ the LSFL with period of 620 nm and HSFL with period around 230-250 nm have been discovered (see Fig. 8). Fig. 8, b, c, demonstrate the beginning of modification of GaSe surface assisted with surface defects. This process has been observed at lower fluence. At Fig. 8, d, e, one can see exfoliated area of GaSe sample. At Fig. 8, f, the fingerprints of typical LIPSS formed at GaSe surface under fs-laser radiation are marked with 1, 2, 3 and corresponding profiles are presented in Fig. 8, i. Periodic structures marked with 1, 3 have similar direction of ripples but different periods. The period of LIPSS labeled by 2 correlates with the value of period of structure marked 3, but the directions are perpendicular each other. To elucidate the origin of such variety of structures more careful experiments should be carried out.

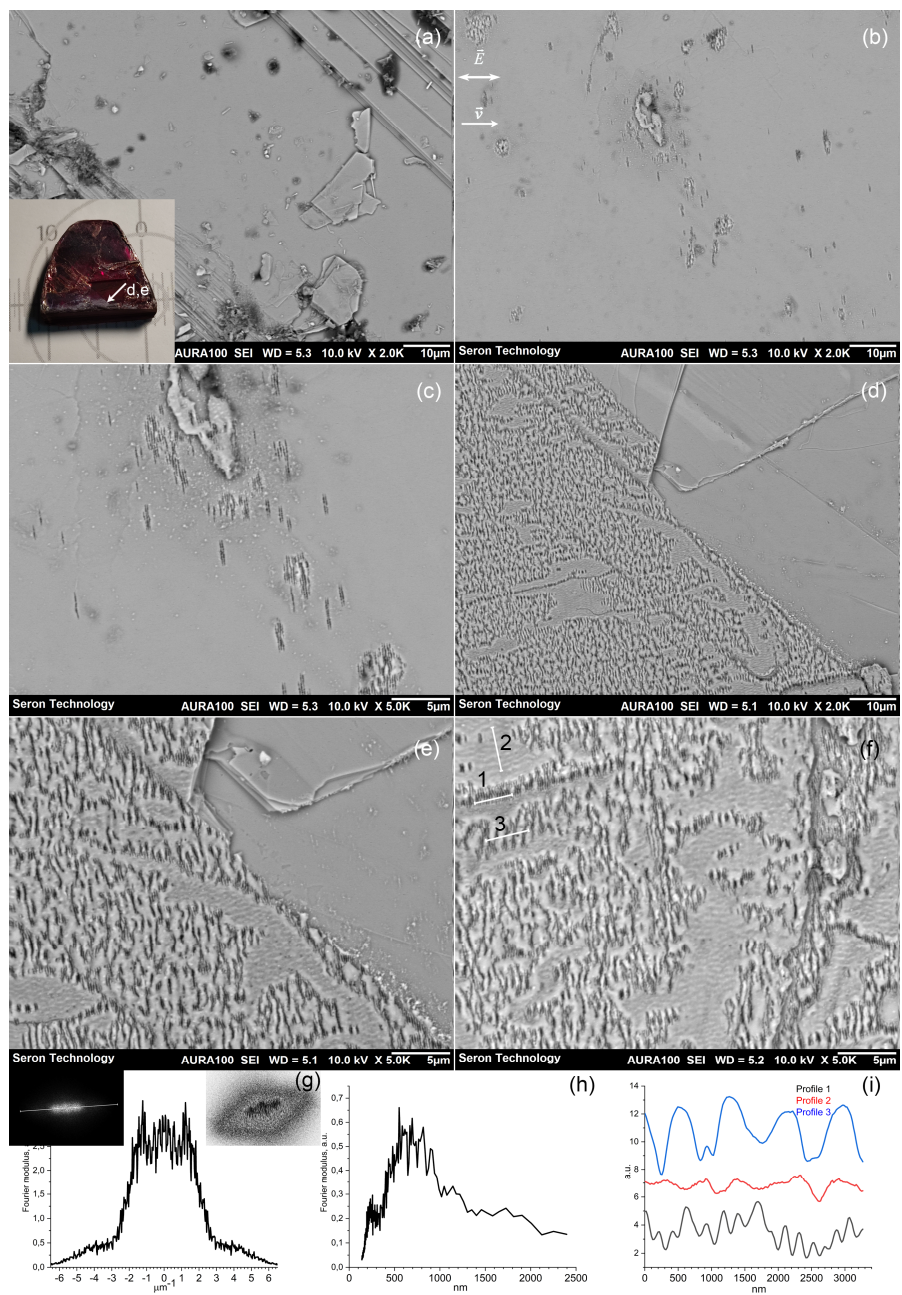

**Fig. 8** (a) SEM image of untreated GaSe surface; (b) SEM image of GaSe surface treated at the pulse energy density of ~0.5 J/cm²; (c) enlarged view of SEM image (b); an area with exfoliated part of the sample (d) and its enlarged view (e); SEM image of treated GaSe and different types of LIPSSs marked with 1, 2, 3 (f); insets in (g) present 2D Fourier transform of SEM image (f), corresponding Fourier transform profile along the line on 2D Fourier transform and corresponding Fourier transform profile in nanometers (g,h), surface profles marked with 1, 2, 3 (i). An inset in (a) presents the view of the GaSe sample

## 5. The femtosecond laser processing of indirect band gap semiconductors

*Indirect band gap semiconductor silicon (Si)* belongs to the most widely used semiconductors for various applications. Laser processing of silicon is actively studied, as Si demonstrate some specific characteristics under irradiation with ultrashort laser pulses [7, 12, 13, 23, 24, 55]. Different mechanisms of the formation of LIPSS at the Si surface have been considered, namely self-organization of LIPSS, the interference of the incident laser radiation with SEWs, including SPP and counterpropagating plasmon waves. Main characteristics of LIPPS depend on the laser processing conditions (laser polarization, frequency, pulse irradiation energy density, angle of incidence and scanning velocity).

In this research, we briefly present the LIPSS formation on silicon surface, paying special attention on the LIPSS morphology. Polished n-doped silicon (100) wafers have been used for LIPSS formation. Micro- and nanostructuring of silicon surfaces have been obtained by irradiation of fs-laser pulses of the wavelength of 800 nm, the pulse energy of about 1.44 mJ, the pulse duration of ~ 150 fs, the repetition rate of 1 kHz. At Fig. 9 one can see typical LIPSS formed at silicon surface. The orientation of the LIPSS is perpendicular to the laser beam polarization. 2D-FFT analysis of SEM image at Fig. 9, a, determine the period of corresponding LSFL of approximately 540 nm. Additional feature in FFT (310 nm) corresponds to the splitting of LSFL ridges. The existence of ripples with the reduced period could be determined by the second harmonic generation of laser radiation at the silicon surface. SEM images analysis reveals the presence of the strip breaks and loops at LSFL. These defects resemble edge dislocations. Fig. 9 b corresponds to the irradiation conditions with bigger fluence. LSFL at Fig. 9, b, is characterised with the period of about 590 nm. At this SEM image the grooves with supra-micron period perpendicular to LSFL are clearly seen.

Fig. 9, (e and f) demonstrate SEM images of silicon surface treated at a half-velocity compared to the processing conditions of silicon presented at SEM image (b). In this case we obtained more prolonged impact of the laser radiation on the viewed area (e, f) that causes the enlarged efficacy of thermal and ablation processes at this area due to the processing conditions. Accidentally we discovered LSFL under the layer that probably consists of ablated products and/or oxides after mechanical remove of upper layer in this part of the fs-laser treated area. Morphologies on Fig. 9 suggest that we are observing structured surface on (a) and a layer of periodically aggregated particles on (b). Whether there is although structured surface beneath them needs further investigation.

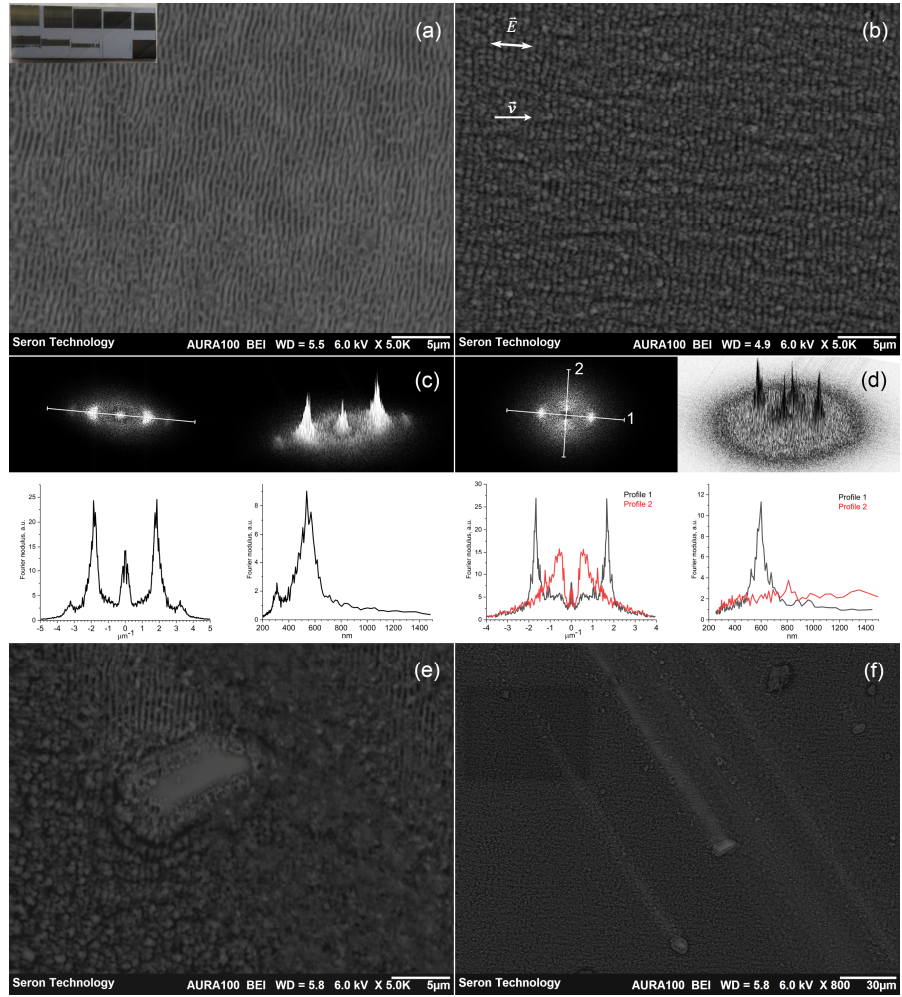

**Fig. 9** SEM image of the fs-laser treated silicon surface (a, b); (e) SEM image of surface of fs-laser treated silicon surface at a half velocity compared to the conditions of SEM image (b); (f) enlarged view of SEM image (e); (c) 2D Fourier transform of a SEM image (a), at the low part of (c) - Fourier transform profile along the line in 2D Fourier transform and corresponding Fourier transform profile in nanometers; (d) 2D Fourier transform of a SEM image (b), at the low part of (d) - Fourier transform profiles along the lines in 2D Fourier transform and corresponding Fourier transform profiles in nanometers. An inset in (a) presents the view of the silicon sample

In our previous paper [56] emission from the processing area of silicon samples has been also studied to clarify the LIPSS formation on silicon. The emission spectra from a moving silicon sample during femtosecond laser treatment revealed wide PL band attributed to the oxidized silicon, second-harmonic generation (SHG) of the laser radiation on the LIPSS and sharp emission lines (EL) of the silicon atoms or ions produced during laser ablation of silicon. The ratio of SHG to PL (or SHG to EL) intensities is suggested as a real-time quality measure for LIPSS generation.

There are few scientific works comparing fs-laser structuring of semiconductors with direct and indirect band gaps [21, 26, 57]. This situation is obviously due to the fact that the resembling kinds of LIPSS are observed in both types of semiconductors, and their nature is considered from more general approaches that take into account the nature of electromagnetic effects or effects of matter reorganization.

Sei T. and co-authors [57] underline the longer relaxation time of the excited carrier in an indirect band gap semiconductor that requires a certain correlation of the positions of not only electrons and holes but also phonons. In the case of fs-laser processed semiconductors such as silicon the interference between the laser beam and the SEW electromagnetic field, as well as the subsequent coupling of the electronic system and the lattice of the solid through electron–phonon coupling, thermal diffusion, electron diffusion and the phase transition of thermal melting have to be considered [26, 27].

# 6. Conclusions

The peculiarities of laser-induced surface structures produced with the femtosecond laser radiation on semiconductors with direct (ZnSe, GaAs, CdZnTe) band gap, with the structurally induced direct-to-indirect band gap transition ($PbI_2$, GaSe) and indirect band gap (Si) have been analysed by scanning electron microscopy, 2D Fourier transform of SEM images, optical spectroscopy. The fs-laser treatment of semiconductors has been performed in the multi-pulse regime in air environment. Under the treatment with the fundamental fs-laser radiation (800 nm, about 130-150 fs) both low spatial frequency LIPSS and high spatial frequency LIPSS have been observed. Other features at the surface (different defects in periodic structure, ablation products, etc.) have been also revealed. In spite of a lot of similar features of formed LIPSSs at semiconductors with direct and indirect band gaps have some factors such as the electron–phonon coupling that could have different effectiveness in each of these semiconductors. Information about laser structuring of some of presented materials in literature is sparse. Thus, further investigations are reasonable.

**Acknowledgments** The authors acknowledge funding from the Ministry of Education and Science of Ukraine (Project No. 19BF051-04, State reg. no. 0119U100300) and appreciate the technical support of the Femtosecond Laser Center for Collective Use of NAS of Ukraine.